\documentclass[a4paper]{jpconf}
\usepackage{graphicx}
\usepackage{cite}

\begin{document}
\title{Data Acquisition for the New Muon $g$-$2$ Experiment at Fermilab}

\author{Wesley Gohn for the Fermilab E989 Collaboration}

\address{Department of Physics and Astronomy, University of Kentucky, Lexington, KY 40506 USA}

\ead{gohn@pa.uky.edu}

\begin{abstract}
A new measurement of the anomalous magnetic moment of the muon, $a_{\mu} \equiv (g-2)/2$, will be performed at the Fermi National Accelerator Laboratory. The most recent measurement, performed at Brookhaven National Laboratory and completed in 2001, shows a 3.3-3.6 standard deviation discrepancy with the Standard Model predictions for $a_\mu$. The new measurement will accumulate 21 times those statistics, measuring $a_\mu$ to 140 ppb and reducing the uncertainty by a factor of 4. 

The data acquisition system for this experiment must have the ability to record deadtime-free records from 700 $\mu$s muon spills at a raw data rate of 18 GB per second. Data will be collected using 1296 channels of $\mu$TCA-based 800 MHz, 12 bit waveform digitizers and processed in a layered array of networked commodity processors with 24 GPUs working in parallel to perform a fast recording and processing of detector signals during the spill. The system will be controlled using the MIDAS data acquisition software package. The described data acquisition system is currently being constructed, and will be fully operational before the start of the experiment in 2017.
\end{abstract}

\section{Introduction}

In the Dirac theory, the muon is a spin 1/2 pointlike particle. It has a magnetic moment given by Eq.~\ref{eq:mu}, 
\begin{equation}
\vec \mu = g\frac{Qe}{2m}\vec s
\label{eq:mu}
\end{equation}
where $g$ was predicted by Dirac to be identically equal to 2~\cite{Dirac:1928hu}. Contributions from QED, weak and hadronic loops, move the Standard Model's predicted value of $g$ very slightly away from 2, so it has become customary to measure the so called muon anomaly $a_\mu=\frac{g-2}{2}$. If a discrepancy with the Standard Model is found, further contributions to $a_\mu$ could come from SUSY, dark photons, or other new physics. Ongoing theoretical and experiment efforts are improving the precision of the Standard Model prediction for $a_\mu$. See for example the first lattice calculation for the light-by-light scattering contribution to the hadronic term~\cite{Blum:2014oka}, a computation of the tenth order QED contribution to $g$-$2$ ($a_\mu(QED) = 116 584 718 951 (80) \times 10^{-14}$)~\cite{Aoyama:2012wk}, and an analysis to reevaluate the hadronic contribution to $g$-$2$ using data from KLOE and BABAR (giving a 3.6$\sigma$ or 2.4$\sigma$ discrepancy from the experimental result depending on the measurement method)~\cite{Davier:2010nc}. 

\begin{figure}[h]
\begin{center}
\includegraphics[width=7cm]{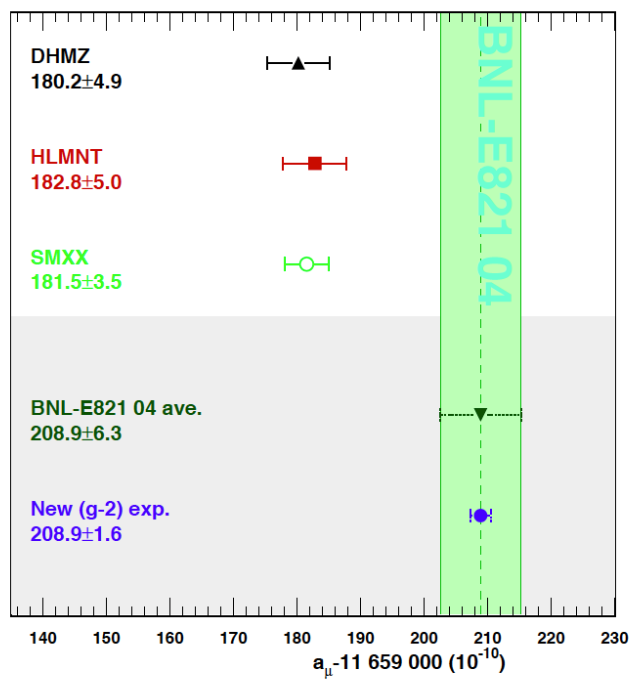}\hspace{2pc}%
\begin{minipage}[b]{14pc}\caption{\label{fig:sigma}Comparison of expected uncertainties on new muon $g$-$2$ measurement to those of BNL-E821 and Standard Model predictions.}
\end{minipage}
\end{center}
\end{figure}

The muon anomaly was measured most recently by Brookhaven National Lab experiment E821~\cite{Bennett:2006fi}, and showed a $\approx3\sigma$ discrepancy with the Standard Model. 
The new muon $g$-$2$ experiment at Fermilab, E989, will measure 21 times the number of muon decays, reducing the uncertainty on this measurement by a factor of four, and plans to begin taking data in early 2017~\cite{Grange:2015fou}. Without theoretical improvements, the discrepancy with the Standard Model~\cite{Davier:2010nc,Hagiwara:2011af} could reach $> 5\sigma$, as shown in Fig.~\ref{fig:sigma}. The increase in data rate required to make this measurement requires the use of a state of the art data acquisition system.

The muon anomaly will be obtained by measuring the precession of muons in a magnetic field. The value of $g$-$2$ will be extracted from precise measurements of $\omega_a$, the difference between
the cyclotron frequency of the muon and the spin precession frequency, and the magnetic field $\vec B$, which are related to $a_\mu$ with the relationship in Eq.~\ref{eq:omega}.

\begin{equation}
\vec\omega_a = -\frac{Qe}{m} a_\mu \vec B 
\label{eq:omega}
\end{equation}

The measurement will be performed using a superconducting magnetic ring that was recently relocated from Brookhaven National Lab in Upton, NY to Fermilab in Batavia, IL. The ring has recently been installed into the new MC-1 building at Fermilab, the steel yokes have all been attached, and cooling of the magnet has begun. The current status of the experiment has been recently reviewed with greater detail in several articles~\cite{Kawall:2013lsa,Lancaster:2014jia,Venanzoni:2014ixa,Grange:2015eea,Kaspar:2015jwa}.

Fills of polarized 3.1 GeV$/c$ muons will be injected into the storage ring at a rate of 12 Hz. The muons will decay into positrons, which will be detected by a suite of calorimeters and tracking detectors in the interior of the ring. Each of the 24 calorimeters is composed of a $9 \times 6$ array of PbF$_2$ crystals with silicon photomultiplier readout~\cite{Fienberg:2014kka}, giving 1296 total channels. The number of muons per 700 $\mu$s fill is expected to be $1.0\times10^4$, but with an estimated acceptance of 10.7\%, we expect to measure $1.1\times10^3$ muons per fill (at an energy $> 1.86$ GeV).

 The calorimeter signals will be digitized using custom $\mu$TCA advanced mezzanine cards (AMCs) built at Cornell University. Each AMC is a 5 channel, 12 bit,  800 MHz waveform digitizer (WFD), the data of which is transmitted through the $\mu$TCA backplane at an expected instantaneous to an AMC13 control card~\cite{Hazen:2013rma} in the $\mu$TCA shelf where the data is aggregated and then read out asynchronously over 10GbE fiber by the data acquisition system. The $\mu$TCA shelf is controlled via IPBus commands sent via a Vadatech MCH. The total data rate from the calorimeters to the DAQ is expected to be 17.4 GB/s given by 1296 channels $\times$ 2 Bytes per sample $\times$ 560000 samples per fill $\times$ 12 fills per second $=$ 17.4 GB/s. 

The system will be calibrated utilizing a laser calibration system that will operate continuously during the data taking of the experiment, as well as for dedicated high rate calibration runs~\cite{Anastasi:2015ssy}. This high trigger rate data must also be accommodated in the data acquisition system.

\section{Data Acquisition System}
The $g$-$2$ DAQ is being built using the MIDAS~\cite{midas} data acquisition software package, which was developed at PSI and widely used at PSI, TRIUMF, and other labs. The design of the DAQ is based largely on prior experience of the collaboration, in particular to the MuLan DAQ~\cite{Tishchenko:2008yk}. The physical system will be composed of a layered array of networked commodity processors with GPUs for parallelization of the data processing. The system must have the ability to provide a deadtime free record of each 700 $\mu$s muon fill at a rate of 12 fills per second with a minimum of 11 ms fill separation, providing a total data rate of 18 GB/s. It will process data from 1296 calorimeter channels, 3 straw tracker stations, and multiple auxiliary detectors. In addition to the derived datasets, the system will have the ability to write raw data for a fraction of the muon fills.

\begin{figure}[h]
\begin{center}
\includegraphics[width=9cm]{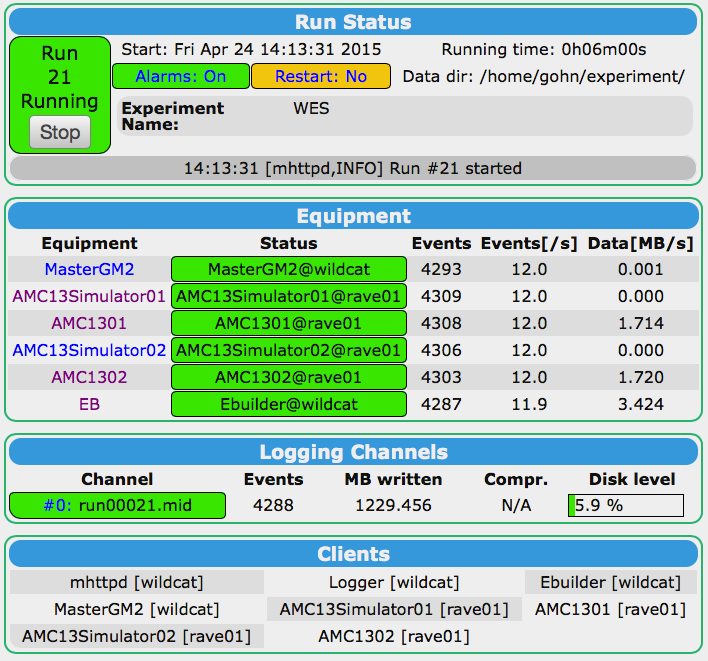}\hspace{2pc}%
\begin{minipage}{12pc}
\caption{\label{fig:midas}The MIDAS web interface showing two calorimeter readout frontends running at full expected data rate.}
\end{minipage}
\end{center}
\end{figure}

MIDAS provides a convenient web interface for control of the experiment, as shown in Fig.~\ref{fig:midas}, as well as the framework for an event builder and data logger, which will output data in a MIDAS binary data format, which can subsequently be processed into ROOT trees and analyzed in the \emph{art} framework that is developed at Fermilab, described in these proceedings by R. Kutschke. 

MIDAS also provides an online database (ODB) used both for saving the configuration of the experiment from run-to-run and also for control of the detectors, as settings that are changed in the ODB are hot-coded to update via the frontend processes in real time. Thus it is possible to change hardware configurations of a WFD component or set the HV on a detector directly via the MIDAS ODB. The values from the ODB can also be stored long term via the MIDAS slow control bus (MSCB), which will write periodic slow control data to a PostgreSQL database.

The frontends for the experiment are written in C/C++. A \emph{master} frontend controls the other frontends via remote procedure calls and synchronizes the data to the muon fill structure, as well as recording a GPS timestamp that will enable subsequent synchronization with the magnetic field data. Additionally a single frontend process will be running to collect data for each detector or detector group. There will be 24 calorimeter frontends, one tracker frontend accumulating data from the three straw tracking detectors, as well as several frontends running to acquire data from the auxiliary detectors such as stored beam monitoring, entrance counters, and electric quadrupole monitors. All of these detectors will be read out via WFDs in $\mu$TCA crates, with the exception of the trackers which are read out via $\mu$TCA-based TDCs, but still controlled via an AMC13 with IPBus. This enables the development of a single frontend code for the acquisition of data from an AMC13 via 10GbE to be similarly utilized for all detectors in the experiment with only a few options for customization between detectors.

The key difference in the processing between the high-rate data coming from the calorimeters and the relatively low-rate data coming from the trackers and auxiliary detectors, is that the calorimeter data will be processed using a hybrid system of multicore CPUs and graphical processing units (GPUs). Data from each of the 24 calorimeters will be processed into multiple derived datasets in a single NVIDIA Tesla K40 GPU. The K40 GPU utilized 2880 cores to provide a vast increase in parallelization over what would be available in traditional CPU-based data processing. The K40 also has a 12GB onboard memory and a memory bandwidth of 288 GB/s. Data is transfered between the system memory and the GPU via PCIe version 3.0, which allows for a significan increase in bandwidth over the the older Tesla K20 GPU and PCIe version 2.0, as shown in Table.~\ref{tab:gpu}.

The CPU multithreading is handled using mutual exclusion (mutex) locks, which help to insure data integrity by dividing the processing for each fill into three distinct threads. A TCP thread is responsible for reading data from the TCP socket, unpacking the data, and copying it to a ring buffer. The GPU thread then performs an asynchronous memory copy to send the data to the GPU, launches the data processing, and the copies the data back to the system memory. The MIDAS thread packs and sends the data into the MIDAS banks and data quality monitoring system.

The GPU multithreading is programmed and optimized using the NVIDIA CUDA libraries. The CUDA code reads the raw digitizer data and processes it into two derived datasets, referred to as the T and Q-methods. The T-method is a standard data taking method in which individual positron hits in the calorimeter are identified. The data is processed into islands of digitizer data where the signal surpasses a given threshold, and the energy of each hit can be measured. The anomalous precession frequency $\omega_a$ is then extracted from a pileup-subtracted time-spectrum. This was the process utilized by the Brookhaven E821 experiment. The main drawback to this method is that pileup causes an early-to-late phase shift in $\omega_a$, which is a significant contribution to the systematic uncertainty. An alternative derived dataset is the so-called Q-method. In the Q-method, individual events are not identified, but instead the detector current is integrated as a function of time, and $\omega_a$ can subsequently be extracted from this time-distribution. This method is attractive because it is hoped to be immune to pileup-related uncertainties. Both of these methods can be implemented simultaneously in a single GPU.

\begin{figure}
\begin{center}
\includegraphics[width=10cm]{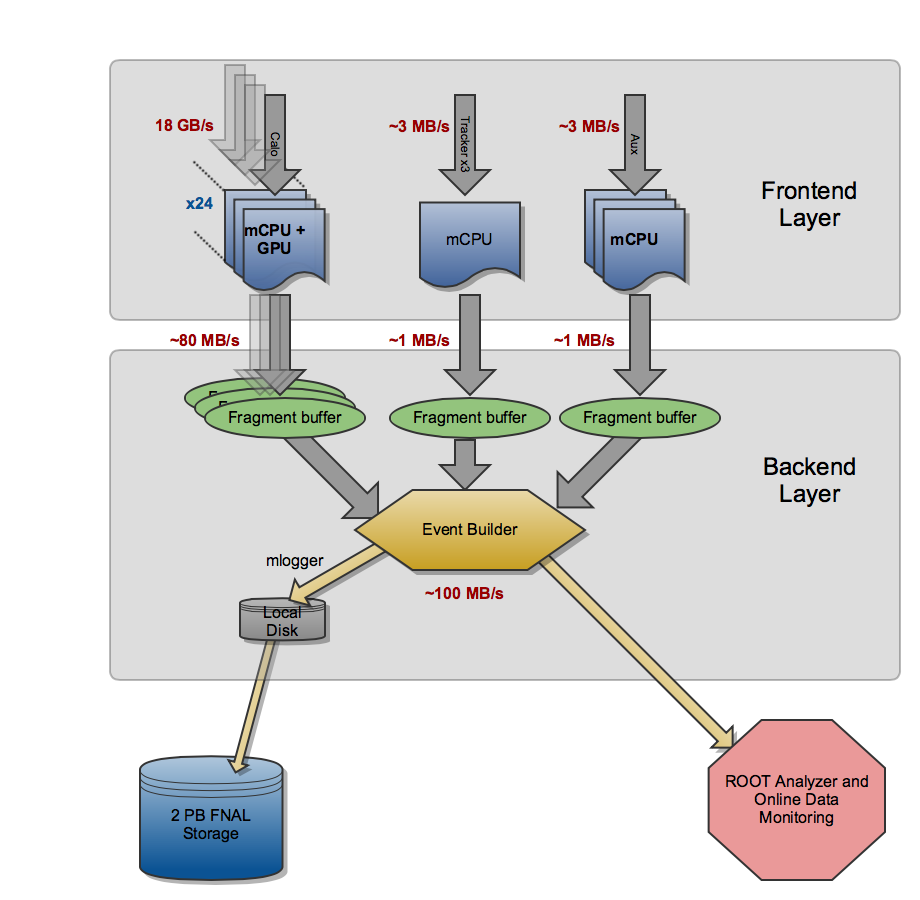}
\caption{\label{fig:flow}The DAQ for $g$-$2$ will be composed of a layered array of processors, with a frontend layer consisting of a hybrid system of GPUs with multi-core CPUs and a backend system used to assemble event fragments and control the experiment.}
\end{center}
\end{figure}

Fig.~\ref{fig:flow} describes the accumulation of data from the frontend processes by the event builder, which will be hosted on a backend machine, and is used to assemble memory fragments from the various frontend processes into a single deadtime free record of each 700 $\mu$s fill. The data will then be copied to the Fermilab central archive via a dedicated 20 Gb/s connection, and also monitored in real time using ROME, a package that interfaces with MIDAS to display ROOT-based analysis in real time for data quality monitoring. The total data output of the experiment is expected to be $\approx 2$ PB.

\begin{table}
\caption{\label{tab:gpu}The GPU data transfer bandwidth was measured with different PCIe versions using Tesla K20 and K40 GPUs. The maximum bandwidth of PCIe version 2.0 is quoted as 500 MB/s per lane and for PCIe version 3.0 is 984.6 MB/s per lane. Tesla GPUs each utilize 16 lanes of PCIe. The memory is copied either as pageable or pinned, which is memory that cannot be swapped.}
\begin{center}
\begin{tabular}{|c|c|c|c|}
\hline
PCIe Version&	GPU	&Host to device, Pageable&	Host to device, Pinned\\
\hline
2&	K20	&3326.6 MB/s&	5028.3 MB/s	\\
3&	K20&	5628.6 MB/s	&6003.6 MB/s	\\
3&	K40	&6647.8 MB/s	&	10044.3 MB/s	\\
\hline
\end{tabular}
\end{center}
\end{table}

\section{Prototyping}

With the start of data taking for the Fermilab muon $g$-$2$ experiment planned for the beginning of 2017, the data acquisition system will be fully constructed and operational by mid-2016. A phased build approach is being utilized. A system comprised of a backend, frontend, gateway, and $\mu$TCA shelf with AMC13 has been assembled and is currently being used for prototyping. The system will be expanded to 25\% of its full strength in mid-2015, and 50\% by the end of 2015. Prototyping is currently underway using simulated data and readout from the AMC13 over 10 GbE, and the full system will be tested with the laser calibration system well before the start of data taking.

The system is currently being prototyped in several test stands. Two test stands for the calorimeter aspect of the DAQ are operating by the University of Kentucky, a test stand for the tracker readout is set up at University College of London, and the data quality monitoring system is being developed at the Joint Institute for Nuclear Research in Dubna. The following will focus primarily on prototyping of the calorimeter DAQ.

The two test stands used for prototyping are essentially the same equipment, but one is set up at the University of Kentucky and the other is set up at Fermilab and operated by the University of Kentucky. Both utilize a single frontend machine and a backend machine connected by 10GbE. Both test stands also currently have a single $\mu$TCA crate with an AMC13 control module and MCH. 

The Fermilab test stand is set up with two machines that are intended to be included in the final data acquisition system. Both computers are using Intel Xeon processors. The backend utilizes an 8-core processor with a 20MB cache, and the frontend uses a 6-core processor with a 12 MB cache. The backend contains 64Gb of 2133 MHz DDR4 ECC RAM, and the frontend has 32 GB of 1600 MHz DDR3. The frontend machine also houses the NVIDIA Tesla K40 GPU. We are experimenting with a setup using two K40 GPUs in one machine, so to test this we are also running the frontend with an additional K20 GPU, but that would be replaced with a K40 if this setup is chosen for running during the experiment.

\begin{figure}[h]
\begin{center}
\includegraphics[width=7cm]{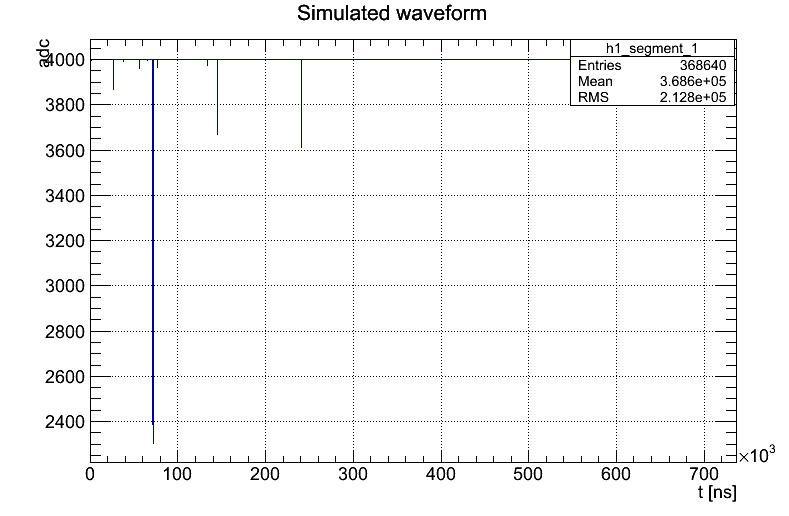}\hspace{2pc}%
\begin{minipage}[b]{14pc}
\caption{\label{fig:sim}Simulated waveform for 700 $\mu$s muon fill as recorded by a 800 MHz, 12 bit waveform digitizer. The pulses represent simulated decays of muons to positrons as seen in the calorimeter.}
\end{minipage}
\end{center}
\end{figure}

Since the WFDs are still under development, much of the DAQ prototyping has utilized a simulator that generates fake data and packages it as would the real AMC13. A data generator makes simulated waveforms for each of the 54 channels of a calorimeter, as shown in Fig.~\ref{fig:sim}. The simulator, which runs as an additional MIDAS frontend called \emph{AMC13Simulator}, then packages the raw waveform data in exactly the same manner as will the AMC13.

The simulated waveforms can then be read and unpacked by the same MIDAS frontend that will read and unpack data from a physical AMC13 with a full compliment of 12 WFDs. Because the simulated waveform has a physical structure analogous to that which we expect from the WFDs, the CUDA algorithms for the T and Q methods can be fully implemented and tested using this method. 

\begin{figure}[h]
\begin{center}
\includegraphics[width=9cm]{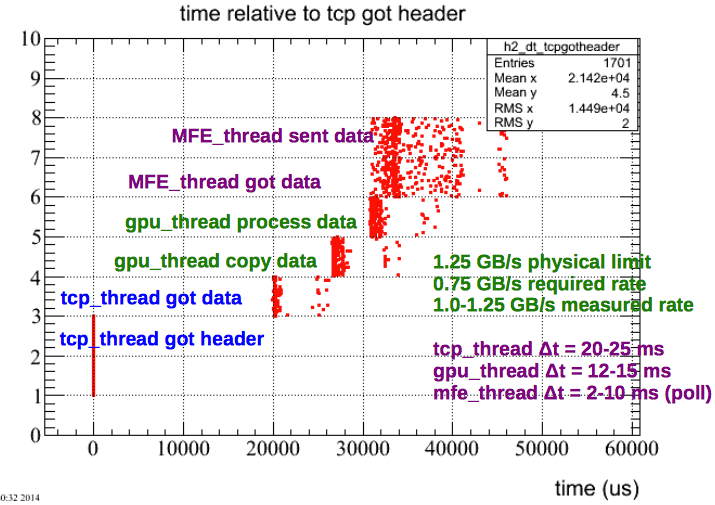}\hspace{2pc}%
\begin{minipage}[b]{14pc}\caption{\label{fig:time}Processing time of data in frontend computer. The vertical axis represents the executions of each subsequent thread, and the horizontal axis provides the time of execution.}
\end{minipage}
\end{center}
\end{figure}

Fig~\ref{fig:time} shows a sample of the processing times recorded for one run with this simulated data. The time stamp of each subsequent process is histogrammed vertically. The \emph{tcp\_thread} reads the data, transfers it to the \emph{GPU\_thread}, which processes the data and sends it on to the \emph{MFE\_thread}. The greatest amount of time is spent reading the data from the TCP socket, followed by copying the data to the GPU memory. Much effort has taken place, and is still underway, to streamline this process as much as possible. Upgrading from PCIe version 2.0 to version 3.0 gave a vast improvement, as was shown earlier in Table.~\ref{tab:gpu}. The size of the data is greatly reduced by the GPU processing, and thus the processes occurring after that happen much faster.

The first readout by the DAQ of the new waveform digitizers that are being developed at Cornell occurred during a test-beam run at SLAC in late 2014, as described in~\cite{Fienberg:2014kka}. Tests were performed using a single-channel WFD, and later a 5-channel WFD that did not yet have the full implementation of it's functionality, but data from the five Rider FPGA's was passed through the $\mu$TCA crate backplane to the AMC13, and read out by the data acquisition system over 10 GbE. We plan to read five of the 5-channel Rider boards by mid-2015, and to have one fully functioning $\mu$TCA shelf (so 12 WFD, and 60 channels) by the Fall of 2015.

\begin{figure}[h]
\begin{center}
\includegraphics[width=10cm]{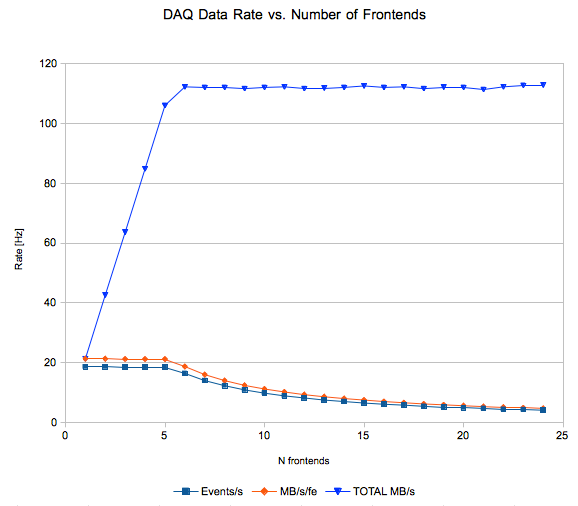}\hspace{2pc}%
\begin{minipage}[b]{12pc}
\caption{\label{fig:rate}Test of event building with 24 frontends. The horizontal axis is the number of frontend processes from 1 to 24. The blue squares represent the rate in events/s, the orange diamonds represent the rate in MB/s for each frontend process, and the blue triangles represent the total data rate.}
\end{minipage}
\end{center}
\end{figure}

\begin{figure}[h]
\begin{center}
\includegraphics[width=10cm]{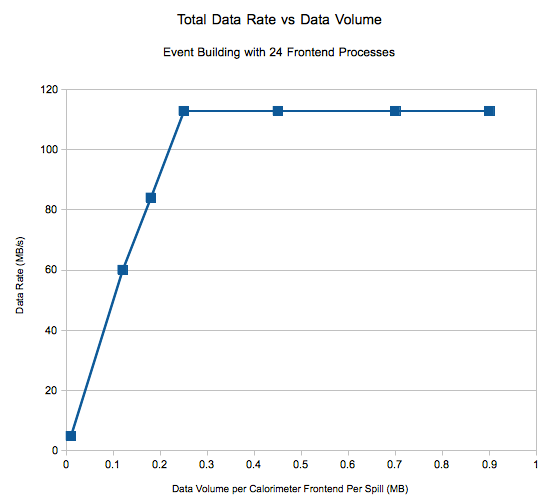}\hspace{2pc}%
\begin{minipage}[b]{12pc}
\caption{\label{fig:eb}Test of event building with 24 frontends running with a varying volume of data.The saturation occurring above 112 MB/s is thought to be a limitation from the speed of memcpys that are being performed in the frontend code.  The event builder comfortably outperforms our projected derived data rate of 80 MB/s.}
\end{minipage}
\end{center}
\end{figure}

A test of the event builder was performed by sending block data from 24 frontends and assembling the event fragments at a full rate of 12 Hz. The \emph{FakeData} frontend has the ability to manually scale the size of the data output, which enabled us to analyze the functionality of the event builder as a function of data size, as shown in Fig.~\ref{fig:rate} and~\ref{fig:eb}. The expected rate of of data that the experiment will write to disk after the GPU processing is expected to be 80-100 MB/s, which the MIDAS event builder outperformed. The test was performed on older, less powerful computers than those that are being set up for the experiment, so we are confident that the event builder will be able to outdo even the displayed performance.

\section{Conclusion}

A data acquisition system is being built for the new Muon $g$-$2$ experiment at Fermilab. The experiment plans to begin data taking in early 2017. The new experiment plans to collect 20 times the statistics of the BNL experiment, which requires a new state-of-the-art acquisition system utilizing parallel data processing in a hybrid system of multi-core CPUs and GPUs. The DAQ will acquire data from 1296 channels of custom $\mu$TCA waveform digitizers, as well as straw trackers and auxiliary detectors at a rate of 18 GB/s. Prototyping of the DAQ is underway, and construction will be complete by mid-2016.

\section*{Acknowledgements}
This research was supported by the National Science Foundation MRI award PHY-1337542. This material is based upon work supported by the U.S. Department of Energy Office of Science, Office of Nuclear Physics under Award Number DE-FG02- 97ER41020.

\section*{References}
\bibliographystyle{h-physrev}
\bibliography{chep}

\begin{thebibliography}{10}

\bibitem{Dirac:1928hu}
P.~A. Dirac,
\newblock Proc.Roy.Soc.Lond. {\bf A117}, 610 (1928).

\bibitem{Blum:2014oka}
T.~Blum, S.~Chowdhury, M.~Hayakawa, and T.~Izubuchi,
\newblock Phys.Rev.Lett. {\bf 114}, 012001 (2015), 1407.2923.

\bibitem{Aoyama:2012wk}
T.~Aoyama, M.~Hayakawa, T.~Kinoshita, and M.~Nio,
\newblock Phys.Rev.Lett. {\bf 109}, 111808 (2012), 1205.5370.

\bibitem{Davier:2010nc}
M.~Davier, A.~Hoecker, B.~Malaescu, and Z.~Zhang,
\newblock Eur.Phys.J. {\bf C71}, 1515 (2011), 1010.4180.

\bibitem{Bennett:2006fi}
Muon g-2, G.~Bennett {\em et~al.},
\newblock Phys.Rev. {\bf D73}, 072003 (2006), hep-ex/0602035.

\bibitem{Grange:2015fou}
Muon g-2, J.~Grange {\em et~al.},
\newblock (2015), 1501.06858.

\bibitem{Hagiwara:2011af}
K.~Hagiwara, R.~Liao, A.~D. Martin, D.~Nomura, and T.~Teubner,
\newblock J. Phys. {\bf G38}, 085003 (2011), 1105.3149.

\bibitem{Kawall:2013lsa}
Muon g-2 New, D.~Kawall,
\newblock AIP Conf.Proc. {\bf 1560}, 106 (2013).

\bibitem{Lancaster:2014jia}
M.~Lancaster,
\newblock PoS {\bf Photon2013}, 041 (2013).

\bibitem{Venanzoni:2014ixa}
Muon g-2, G.~Venanzoni,
\newblock (2014), 1411.2555.

\bibitem{Grange:2015eea}
Muon g-2, J.~Grange,
\newblock (2015), 1501.03040.

\bibitem{Kaspar:2015jwa}
J.~Kaspar,
\newblock (2015), 1504.01201.

\bibitem{Fienberg:2014kka}
A.~Fienberg {\em et~al.},
\newblock Nucl.Instrum.Meth. {\bf A783}, 12 (2015), 1412.5525.

\bibitem{Hazen:2013rma}
E.~Hazen {\em et~al.},
\newblock JINST {\bf 8}, C12036 (2013).

\bibitem{Anastasi:2015ssy}
Muon g-2, A.~Anastasi {\em et~al.},
\newblock Nucl.Instrum.Meth. {\bf A788}, 43 (2015), 1504.00132.

\bibitem{midas}
{http://midas.triumf.ca},
\newblock TRIUMF MIDAS homepage, 2015.

\bibitem{Tishchenko:2008yk}
V.~Tishchenko {\em et~al.},
\newblock Nucl.Instrum.Meth. {\bf A592}, 114 (2008), 0802.1029.

\end{thebibliography}
\end{document}